\documentclass[american,aps,pra,reprint,superscriptaddress,showpacs]{revtex4-1}
\usepackage[T1]{fontenc}
\usepackage[latin9]{inputenc}
\usepackage{babel}
\usepackage{amsthm}
\usepackage{amsmath}
\usepackage{amssymb}
\usepackage{graphicx}
\usepackage[unicode=true,pdfusetitle,
 bookmarks=true,bookmarksnumbered=false,bookmarksopen=false,
 breaklinks=false,pdfborder={0 0 0},backref=false,colorlinks=false]
 {hyperref}
\hypersetup{
 colorlinks,linkcolor=myurlcolor,citecolor=myurlcolor,urlcolor=myurlcolor}

\makeatletter
\@ifundefined{textcolor}{}
{%
 \definecolor{BLACK}{gray}{0}
 \definecolor{WHITE}{gray}{1}
 \definecolor{RED}{rgb}{1,0,0}
 \definecolor{GREEN}{rgb}{0,1,0}
 \definecolor{BLUE}{rgb}{0,0,1}
 \definecolor{CYAN}{cmyk}{1,0,0,0}
 \definecolor{MAGENTA}{cmyk}{0,1,0,0}
 \definecolor{YELLOW}{cmyk}{0,0,1,0}
 }
\theoremstyle{plain}
\newtheorem{thm}{\protect\theoremname}

\usepackage{babel}
\usepackage{babel}
\usepackage{times}\usepackage{txfonts}\usepackage{braket}\usepackage{colortbl}\definecolor{myurlcolor}{rgb}{0,0,0.7}

\providecommand{\theoremname}{Theorem}

\providecommand{\theoremname}{Theorem}

\makeatother

\providecommand{\theoremname}{Theorem}

\begin{document}

\title{Concentrating tripartite quantum information}

\author{Alexander Streltsov}

\affiliation{ICFO-The Institute of Photonic Sciences, Mediterranean Technology
Park, 08860 Castelldefels (Barcelona), Spain}

\author{Soojoon Lee}

\affiliation{Department of Mathematics and Research Institute for Basic Sciences,
Kyung Hee University, Seoul 130-701, Korea}

\author{Gerardo Adesso}

\affiliation{$\mbox{School of Mathematical Sciences, The University of Nottingham, University Park, Nottingham NG7 2RD, United Kingdom}$}
\begin{abstract}
We introduce the concentrated information of tripartite quantum states.
For three parties Alice, Bob, and Charlie, it is defined as the maximal
mutual information achievable between Alice and Charlie via local
operations and classical communication performed by Charlie and Bob.
We derive upper and lower bounds to the concentrated information,
and obtain a closed expression for it on several classes of states
including arbitrary pure tripartite states in the asymptotic setting.
We show that distillable entanglement, entanglement of assistance,
and quantum discord can all be expressed in terms of the concentrated
information, thus revealing its role as a unifying informational primitive.
We finally investigate quantum state merging of mixed states with
and without additional entanglement. The gap between classical and
quantum concentrated information is proven to be an operational figure
of merit for mixed state merging in absence of additional entanglement.
Contrary to pure state merging, our analysis shows that classical
communication in both directions can provide advantage for merging
of mixed states.
\end{abstract}

\pacs{03.67.Hk, 03.65.Ud, 89.70.Cf}

\date{June 30, 2015}

\maketitle
\emph{Introduction}.---Correlations between parts of a composite system
are crucial to dictate its collective behavior and to determine its
usefulness for functional tasks involving the correlated components.
This is true both for physical models in condensed matter and statistical
mechanics \cite{Wolf2008}, and for complex systems in the biological,
engineered and social domains \cite{Shalizi}. In classical and quantum
systems, correlations between two parties are generally quantified
by the \textit{mutual information}. In a thermodynamic context, mutual
information quantifies the amount of work required to erase all the
correlations established between two parties \cite{Groisman2005}.
In the context of quantum communication \cite{Wilde}, mutual information
plays a fundamental role to describe the classical capacity of a noisy
quantum channel connecting the two parties \cite{Bennett1999}. Maximizing
the mutual information between, say, Alice and Charlie, within a larger
system potentially involving other cooperative or competitive players,
ensures that a reliable communication channel is established between
the chosen sender and receiver, so that Alice and Charlie can implement
quantum cryptography or quantum state transfer protocols with high
success \cite{Nielsen2000}.

In this Letter we introduce and study a quantum informational task
that we name \textit{information concentration}. We consider a general
communication scenario involving three parties, Alice, Bob, and Charlie,
who initially share an arbitrary mixed quantum state. Our main question
can then be formulated as follows:\emph{``How much can Charlie learn
about Alice by asking Bob?''} To answer this question we analyze
the task of maximizing the mutual information between Alice and Charlie
via a cooperative strategy by Charlie and Bob only relying on local
operations and classical communication (LOCC). The corresponding maximal
mutual information between Alice and Charlie is termed \textit{concentrated
information (CI)}.

\begin{figure}[t]
\includegraphics[width=0.7\columnwidth]{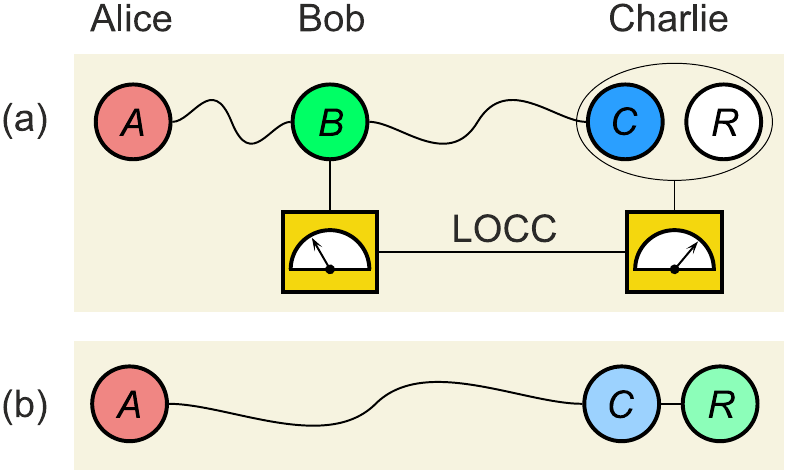} \caption{\label{fig:1}(color online). Concentrating information in tripartite
quantum states. Panel (a) shows the initial situation: Alice, Bob,
and Charlie share a quantum state $\rho^{ABC}$; additionally, Charlie
has access to a quantum register $R$. Bob and Charlie perform local
operations and classical communication (LOCC), aiming to maximize
the mutual information between Alice and Charlie. The final state
shared by Alice and Charlie is illustrated in Panel (b).}
\end{figure}

In the classical domain this quantity coincides with the total mutual
information between Alice and the remaining two parties, since in
this case Bob can share all his knowledge with Charlie via a classical
channel. However, the situation changes completely if quantum theory
is applied. As we will show, the CI is in general below the maximal
value achievable in the classical case. We derive upper and lower
bounds to the CI which depend on classical and quantum correlations
in different partitions of the original tripartite state. Remarkably,
when the three players share asymptotically many copies of an arbitrary
pure state, we obtain a closed expression for the CI, only depending
on the initial entropic degrees of Alice's and Charlie's subsystems.
The CI can be further evaluated exactly in some classes of mixed states.
The broad relevance of the concept is underlined by showing that distillable
entanglement \cite{Bennett1996}, entanglement of assistance \cite{DiVincenzo1999},
and quantum discord \cite{Zurek2000,Ollivier2001} can all be expressed
in general as exact functions of CI.

Finally, we study the usefulness of the CI in the context of quantum
state merging \cite{Horodecki2005,Horodecki2007}. We extend state
merging to the realistic case of mixed states, and show that for this
generalized task classical communication in one direction is strictly
less powerful than general LOCC. Furthermore, by exploiting recent
breakthrough results on conditional mutual information \cite{Fawzi2014,Sutter2015},
we prove that the CI yields a faithful figure of merit for \emph{LOCC
quantum state merging} \emph{(LQSM)}, a variant of state merging operating
on mixed states without additional entanglement. The results of this
Letter provide fundamental and practical advances for quantum information
theory and its applications in a multipartite scenario.

\emph{Concentrated information: Setting and definitions}.---We consider
three parties, Alice, Bob, and Charlie, sharing a quantum state $\rho=\rho^{ABC}$.
The aim of Bob and Charlie is to concentrate their mutual information
with Alice on Charlie's side via LOCC. To this aim, Charlie makes
use of an auxiliary quantum register $R$, so that the total initial
state is given by 
\begin{equation}
\sigma_{i}=\rho^{ABC}\otimes\rho^{R}.\label{eq:initial}
\end{equation}

In the concentration process, Bob and Charlie perform an LOCC protocol
which maximizes the mutual information between Alice and Charlie (see
Fig.~\ref{fig:1}). Noting that the total system of Charlie consists
of two subsystems $C$ and $R$, the maximal mutual information achievable
in this process is given by 
\begin{align}
{\cal I}(\rho) & =\sup_{\Lambda}I^{A:CR}(\sigma_{f}).\label{eq:I}
\end{align}
In the above expression, $I^{A:CR}$ is the mutual information between
Alice's system $A$ and Charlie's system $CR$, the supremum is taken
over all LOCC protocols $\Lambda=\Lambda_{B\leftrightarrow CR}$ between
Bob and Charlie, and the final state $\sigma_{f}=\sigma_{f}^{ACR}$
is the state shared by Alice and Charlie after the application of
the LOCC protocol $\Lambda$ on the initial state $\sigma_{i}$: 
\begin{equation}
\sigma_{f}=\mathrm{Tr}_{B}[\Lambda[\sigma_{i}]].\label{eq:final}
\end{equation}
The quantity defined in Eq.~(\ref{eq:I}) will be referred to as
\emph{concentrated information (CI)}. We will also consider the case
of one-way LOCC where the classical communication is directed from
Bob to Charlie only. The maximal mutual information in this case will
be called \emph{one-way concentrated information}, and we will denote
it by ${\cal I}_{\rightarrow}$. We will also study the situation
where a large number of copies of the state $\rho$ is available.
The corresponding regularized CI is given as 
\begin{equation}
{\cal I}^{\infty}(\rho)=\lim_{n\rightarrow\infty}\frac{1}{n}{\cal I}(\rho^{\otimes n}),
\end{equation}
and its one-way version will be denoted by ${\cal I}_{\rightarrow}^{\infty}$.

At this point it is useful to note that the CI is never smaller than
its one-way version ${\cal I}_{\rightarrow}$ and never larger than
the total mutual information $I^{A:BC}$: 
\begin{equation}
{\cal I}_{\rightarrow}(\rho)\leq{\cal I}(\rho)\leq I^{A:BC}(\rho).\label{eq:inequality}
\end{equation}
The first inequality is evident by observing that one-way LOCC is
a restricted version of general LOCC. The second inequality represents
the fact that Bob and Charlie cannot concentrate more mutual information
than is initially present in the total state $\rho$. The proof follows
by noting that any operation acting on the systems of Bob and Charlie
cannot increase their mutual information with Alice \cite{Vedral2002},
and thus: $I^{A:CR}(\sigma_{f})\leq I^{A:BCR}(\sigma_{i})$. Together
with the fact that $I^{A:BCR}(\sigma_{i})=I^{A:BC}(\rho)$ this completes
the proof of Eq.~(\ref{eq:inequality}). We remark that the CI has
a natural interpretation as the amount of information Charlie can
obtain about Alice by asking Bob, within the considered quantum communication
scenario.

\emph{Bounding the CI}.---Having introduced the CI, we will now show
a powerful upper bound, which also relates ${\cal I}$ to the distillable
entanglement $E_{d}$. As we will also see below in this Letter, the
bound is tight in a large number of relevant scenarios, including
all pure states in the asymptotic limit. 
\begin{thm}
\label{thm:1}CI is bounded above as follows: 
\begin{equation}
{\cal I}(\rho)\leq\min\left\{ I^{A:BC}(\rho),S(\rho^{A})+E_{d}^{AB:C}(\rho)\right\} .\label{eq:upbound}
\end{equation}
\par 
\end{thm}
We note that the same bound also applies to the regularized concentrated
information ${\cal I}^{\infty}$. The proof of the theorem can be
found in the Supplemental Material \cite{epaps}, see section 1 there.

\noindent Due to Eq.~(\ref{eq:inequality}), the above theorem also
provides an upper bound on the one-way CI. Similarly, any lower bound
on ${\cal I}_{\rightarrow}$ is also a lower bound for ${\cal I}$.
We will now show that ${\cal I}_{\rightarrow}$ can be bounded below
as follows: 
\begin{equation}
{\cal I}_{\rightarrow}(\rho)\geq\max\left\{ I^{A:C}(\rho^{AC}),I^{A:B}(\rho^{AB})-\delta^{A|B}(\rho^{AB})\right\} ,\label{eq:lower bound}
\end{equation}
where $\delta$ is the quantum discord \cite{Zurek2000,Ollivier2001},
a measure of quantumness of correlations (for more details and alternative
definitions see also \cite{Henderson2001,Streltsov2011,Piani2011,Modi2012,Brandao2013,Streltsov2014}).
The inequality ${\cal I}_{\rightarrow}(\rho)\geq I^{A:C}(\rho^{AC})$
can be seen by noting that this amount of mutual information between
Alice and Charlie is always achieved if Bob and Charlie do not interact.
On the other hand, the inequality ${\cal I}_{\rightarrow}(\rho)\geq I^{A:B}(\rho^{AB})-\delta^{A|B}(\rho^{AB})$
can be seen by noting that erasing Charlie's system cannot increase
the CI: ${\cal I}_{\rightarrow}(\rho)\geq{\cal I}_{\rightarrow}(\rho^{AB}\otimes\ket{0}\bra{0}^{C})$.
To complete the proof of Eq.~(\ref{eq:lower bound}), we note that
for states of the form $\rho^{AB}\otimes\rho^{C}$ the CI and its
one-way version coincide, and are given by \cite{Streltsov2013,Streltsov2014}
\begin{equation}
{\cal I}(\rho^{AB}\otimes\rho^{C})={\cal I}_{\rightarrow}(\rho^{AB}\otimes\rho^{C})=I^{A:B}(\rho^{AB})-\delta^{A|B}(\rho^{AB}).\label{eq:discord}
\end{equation}

We note that both of the aforementioned quantities bounding the CI,
namely the distillable entanglement and the quantum discord, are usually
difficult to compute for an arbitrary state. However, closed expressions
for both quantities are known for many important families of states
\cite{Horodecki2009,Modi2012}. For instance, the distillable entanglement
can be evaluated exactly for all maximally correlated states, and
that quantum discord can be evaluated for any state $\rho^{AB}$ of
rank two, if the subsystem $A$ is a qubit. This renders the bounds
on the CI analytically accessible in several relevant cases. Finally,
we also mention that the bounds provided in Eqs.~(\ref{eq:upbound})
and (\ref{eq:lower bound}) can be adapted to obtain alternative upper
and lower bounds on the CI, which may be easier to evaluate. In particular,
any upper bound on the distillable entanglement (such as the logarithmic
negativity \cite{vidalwerner,eisertphd,plenioprl}, which is a computable
entanglement monotone related to the entanglement cost under operations
preserving the positivity of the partial transpose \cite{eisertppt,ishi})
provides a (looser) upper bound on the CI via Eq.~(\ref{eq:upbound}).
Similarly, (looser) lower bounds can be derived from Eq.~(\ref{eq:lower bound})
by providing upper bounds on quantum discord; since quantum discord
is defined as a minimization problem, it is easy to provide computable
bounds also in this situation; see e.g. \cite{Girolami2011,HuFan2013}.

\emph{Exact evaluation of CI}.---We now show that, impressively, closed
formulae for the CI can be obtained for a number of relevant classes
of states. We start by considering the situation where Alice, Bob,
and Charlie share a pure state $\ket{\psi}=\ket{\psi}^{ABC}$. In
this case, the one-way CI is given exactly by 
\begin{equation}
{\cal I}_{\rightarrow}(\ket{\psi})=S(\rho^{A})+E_{a}(\rho^{AC}).\label{eq:pure states}
\end{equation}
Here, $E_{a}$ is the entanglement of assistance which was defined
in \cite{DiVincenzo1999} as follows: $E_{a}(\rho^{AC})=\max\sum_{i}p_{i}E_{d}(\ket{\psi_{i}}^{AC})$.
The maximum is taken over all decompositions of the state $\rho^{AC}$,
while the distillable entanglement of a pure state $\ket{\psi_{i}^{AC}}$
is equal to the von Neumann entropy of the reduced state \cite{Bennett1996}:
$E_{d}(\ket{\psi_{i}}^{AC})=S(\rho_{i}^{A})$. For the proof of Eq.~(\ref{eq:pure states})
we refer to the Supplemental Material \cite{epaps}, see section 2
there. We will now evaluate the regularized CI for an arbitrary tripartite
pure state $\ket{\psi}=\ket{\psi}^{ABC}$. Remarkably, in this scenario
${\cal I}^{\infty}$ and ${\cal I}_{\rightarrow}^{\infty}$ both coincide
with the bound provided in Theorem \ref{thm:1}, i.e., the bound is
tight for all pure states in the asymptotic setting. 
\begin{thm}
\label{thm:2} For any pure state $\ket{\psi}=\ket{\psi}^{ABC}$ it
holds: 
\begin{equation}
{\cal I}^{\infty}(\ket{\psi})={\cal I}_{\rightarrow}^{\infty}(\ket{\psi})=S(\rho^{A})+\min\{S(\rho^{A}),S(\rho^{C})\}.
\end{equation}

\end{thm}
This theorem provides a simple expression for the regularized CI of
pure states, and shows that one-way LOCC operations suffice for optimal
information concentration in the asymptotic setting. For the proof
of the theorem we refer to the Supplemental Material \cite{epaps},
see section 3 there.

Finally, we consider an instance of mixed states, where Bob is in
possession of two particles $B_{1}$ and $B_{2}$, each of them being
correlated exclusively with Alice or Charlie. If the state shared
by Alice and Bob is pure, the scenario is covered by states of the
form 
\begin{equation}
\rho=\ket{\psi}\bra{\psi}^{AB_{1}}\otimes\rho^{B_{2}C}.\label{eq:product}
\end{equation}
As we show in section 4 of the Supplemental Material \cite{epaps},
the results presented in this Letter allow to evaluate the regularized
CI for this set of states: 
\begin{equation}
\mathcal{I}^{\infty}(\rho)=S(\rho^{A})+\min\{S(\rho^{A}),E_{d}(\rho^{B_{2}C})\}.\label{eq:distillable_entanglement}
\end{equation}
Importantly, this implies that the bound provided in Theorem \ref{thm:1}
is asymptotically saturated for all states given in Eq.~(\ref{eq:product}).

\emph{CI as a unifying quantum informational primitive.}---The approach
presented in this Letter allows to unify three fundamental quantities
in quantum information theory: distillable entanglement $E_{d}$ \cite{Bennett1996},
entanglement of assistance $E_{a}$ \cite{DiVincenzo1999}, and quantum
discord $\delta$ \cite{Zurek2000,Ollivier2001}. As we will see in
the following, all these quantities can be traced to a common origin,
since all of them can be written in terms of the CI.

For $E_{a}$ this can be seen by using Eq.~(\ref{eq:pure states}),
which implies that the entanglement of assistance of a state $\rho^{AC}$
is related to the one-way CI as follows: $E_{a}(\rho^{AC})={\cal I}_{\rightarrow}(\ket{\psi})-S(\rho^{A})$,
where $\ket{\psi}=\ket{\psi}^{ABC}$ is a purification of $\rho^{AC}$.
The relation to quantum discord $\delta$ is evident from Eq.~(\ref{eq:discord}),
according to which the amount of discord in a state $\rho^{AB}$ can
be expressed in terms of CI as follows: $\delta^{A|B}(\rho^{AB})=I^{A:B}(\rho^{AB})-\mathcal{I}(\rho^{AB}\otimes\rho^{C})$,
where $\rho^{C}$ is an arbitrary state of Charlie's system $C$.
Finally, the relation between CI and distillable entanglement is given
by Eq.~(\ref{eq:distillable_entanglement}), which implies that $E_{d}(\rho^{B_{2}C})=\mathcal{I}^{\infty}(\rho)-\log_{2}d_{A}$,
for an arbitrary state $\rho^{B_{2}C}$, with $\rho=\ket{\phi^{+}}\bra{\phi^{+}}^{AB_{1}}\otimes\rho^{B_{2}C}$,
$\ket{\phi^{+}}^{AB_{1}}=\sum_{i}\ket{ii}^{AB_{1}}/\sqrt{d_{A}}$,
and $d_{A}=d_{B_{1}}=d_{B_{2}}$.

It is straightforward to extend the aforementioned results to entanglement
of formation $E_{f}$ and entanglement cost $E_{c}$ by using the
relation between quantum discord and entanglement of formation \cite{Koashi2004,Fanchini2011,Madhok2011,Cavalcanti2011},
and recalling that $E_{c}$ is equal to regularized $E_{f}$ \cite{Hayden2001}.
It is thus reasonable to expect that other important quantities might
be also recast in terms of the CI.

\emph{LOCC quantum state merging (LQSM)}.---We will now show that
the task of concentrating information presented in this Letter is
closely related to the task of merging quantum states via LOCC, that
we analyze here. In the latter task, Bob and Charlie aim to merge
their parts of the total state $\rho=\rho^{ABC}$ on Charlie's side
via LOCC, while preserving the coherence with Alice. To this end,
Charlie has access to an additional register $R$, and the overall
initial state $\sigma_{i}$ is again given by Eq.~(\ref{eq:initial}).
It is instrumental to compare this task to the standard quantum state
merging as presented in \cite{Horodecki2005,Horodecki2007}. In contrast
to that well established protocol, in LQSM Bob and Charlie are not
allowed to use any additional entangled resource states, and the overall
state $\rho$ is not restricted to be pure.

We now introduce the fidelity of LQSM as follows: 
\begin{equation}
{\cal F}(\rho)=\sup_{\Lambda}F(\sigma_{f},\sigma_{t})\label{eq:F}
\end{equation}
with Uhlmann fidelity $F(\rho,\sigma)=\mathrm{Tr}(\sqrt{\rho}\sigma\sqrt{\rho})^{1/2}$.
Here, the desired target state $\sigma_{t}=\sigma_{t}^{ACR}$ is the
same state as $\rho=\rho^{ABC}$ up to relabeling the systems $B$
and $R$. The final state $\sigma_{f}$ was already introduced in
Eq.~(\ref{eq:final}), and the supremum is taken over all LOCC operations
$\Lambda=\Lambda_{B\leftrightarrow CR}$ between Bob and Charlie.

The relevance of the quantity defined in Eq.~(\ref{eq:F}) comes
from the fact that it faithfully captures the performance of the considered
task. In particular, a state $\rho$ admits \emph{perfect LQSM} if
and only if ${\cal F}(\rho)=1$, while ${\cal F}(\rho)<1$ otherwise.
As we will see in a moment, the fidelity is closely related to the
gap between quantum and classical CI, which can then be regarded as
a faithful figure of merit for LQSM on its own right. In particular,
we will find that perfect LQSM is possible if and only if the CI is
equal to the total mutual information $I^{A:BC}$: 
\begin{equation}
{\cal F}(\rho)=1\Leftrightarrow{\cal I}(\rho)=I^{A:BC}(\rho),\label{eq:perfectLQSM}
\end{equation}
while ${\cal I}(\rho)<I^{A:BC}(\rho)$ otherwise. This result implies
an operational equivalence between information concentration and LQSM:
a state admits perfect LQSM if and only if it admits perfect information
concentration, i.e., if all the mutual information available in the
state can be concentrated on Charlie's side. To prove the statement
in Eq.~(\ref{eq:perfectLQSM}) we will establish a link between ${\cal F}$
and ${\cal I}$ formalized by the following theorem. \begin{thm}
\label{thm:3}The fidelity of LQSM is bounded below as 
\begin{equation}
{\cal F}(\rho)\geq2^{-\frac{1}{2}[I^{A:BC}(\rho)-{\cal I}(\rho)]}.
\end{equation}
\par \end{thm} The proof of the theorem is based on very recent
results from \cite{Fawzi2014} and can be found in the Supplemental
Material \cite{epaps}, see section 5 there. From this result it is
evident that perfect information concentration implies perfect LQSM:
${\cal I}(\rho)=I^{A:BC}(\rho)\Rightarrow{\cal F}(\rho)=1$. The other
direction follows straightforwardly by continuity of the mutual information.

These results demonstrate that the gap $I^{A:BC}(\rho)-{\cal I}(\rho)$
has an inherent operational meaning, quantifying the deviation from
perfect LQSM. Note that this gap is a genuinely quantum feature, and
vanishes for fully classical states. In the classical domain all the
mutual information available in the state can be concentrated via
classical communication.

Furthermore, we will provide another necessary condition for perfect
LQSM. In particular, Bob and Charlie can perfectly merge their systems
via LOCC on Charlie's side only if the state $\rho=\rho^{ABC}$ satisfies
the inequality $E^{AB:C}(\rho)\geq E^{A:BC}(\rho)$ for all entanglement
measures $E$. The statement can be proven directly by using the fact
that any valid entanglement monotone $E$ cannot increase under LOCC
\cite{Horodecki2009}. This implies that any state $\rho$ which violates
the above inequality for some entanglement measure does not allow
for perfect LQSM.

\emph{Quantum state merging of mixed states}.---Finally, we will show
that the novel concepts of information concentration and LQSM are
also useful in the context of conventional quantum state merging \cite{Horodecki2005,Horodecki2007}.
We first introduce the asymptotic fidelity of LQSM, ${\cal F}^{\infty}(\rho)=\lim_{n\rightarrow\infty}{\cal F}(\rho^{\otimes n})$,
and note that a state $\rho$ allows for \emph{perfect asymptotic
LQSM} if and only if its asymptotic fidelity is ${\cal F}^{\infty}(\rho)=1$.
As we show in section 6 of the Supplemental Material \cite{epaps},
perfect asymptotic LQSM implies perfect asymptotic information concentration:
\begin{equation}
{\cal F}^{\infty}(\rho)=1\Rightarrow{\cal I}^{\infty}(\rho)=I^{A:BC}(\rho).\label{eq:asymptotic}
\end{equation}
This result means that Bob and Charlie cannot merge their state via
LOCC even in the asymptotic scenario, if the regularized CI is below
the total mutual information $I^{A:BC}$. The importance of this result
lies in the fact that the regularized CI can be evaluated exactly
in a large number of relevant scenarios, as was demonstrated previously
in this Letter.

Using the tools presented above, we are now in position to extend
quantum state merging to mixed states in the following way. For a
given state $\rho=\rho^{ABC}$, we supplement Bob and Charlie with
additional entangled states $\ket{\phi}=\ket{\phi}^{B'C'}$. If we
now adjust these states such that the CI of the total state $\rho\otimes\ket{\phi}\bra{\phi}$
becomes equal to the total mutual information $I^{A:BC}$, the amount
of entanglement in $\ket{\phi}$ provides a lower bound on the amount
of resources needed to merge the mixed state $\rho$. This shows how
the tools just developed can be used to gain new results in the established
framework of state merging.

In the next step we will demonstrate how results from quantum state
merging can be carried over to LQSM. In particular, the results presented
in \cite{Horodecki2005,Horodecki2007} imply that Bob and Charlie
can asymptotically merge their parts of a pure state $\ket{\psi}=\ket{\psi}^{ABC}$
via LOCC if and only if their conditional entropy $S(\rho^{BC})-S(\rho^{C})$
is not positive. This result can be immediately extended to mixed
states: if a state $\rho=\rho^{ABC}$ has nonpositive conditional
entropy, it allows for perfect LQSM asymptotically, i.e., $S(\rho^{BC})-S(\rho^{C})\leq0\Rightarrow{\cal F}^{\infty}(\rho)=1.$
Together with Eq.~(\ref{eq:asymptotic}) this means that perfect
asymptotic information concentration is also possible in this case.
Note that the converse is not true in general: there exist mixed states
$\rho$ which allow for perfect LQSM, but have positive conditional
entropy.

Finally, we will show that, in quantum state merging of mixed states,
general LOCC are strictly more powerful than one-way LOCC. This is
notable, since both procedures are instead equivalent in the traditional
quantum state merging of pure states \cite{Horodecki2005,Horodecki2007},
for which classical communication in both directions does not provide
any advantage. In particular, we will present a family of states allowing
for perfect state merging with general LOCC in the single-shot scenario,
but which cannot be merged via one-way LOCC even asymptotically. The
following family of states has this property: 
\begin{align}
\rho & =\frac{1}{4}\left(\ket{0}\bra{0}^{B}\otimes\ket{00}\bra{00}^{AC}+\ket{1}\bra{1}^{B}\otimes\ket{10}\bra{10}^{AC}\right.\\
 & \left.+\ket{\psi}\bra{\psi}^{B}\otimes\ket{01}\bra{01}^{AC}+\ket{\psi_{\bot}}\bra{\psi_{\bot}}^{B}\otimes\ket{11}\bra{11}^{AC}\right),\nonumber 
\end{align}
with mutually orthogonal states $\ket{\psi}$ and $\ket{\psi_{\bot}}$
such that $0<|\braket{0|\psi}|<1$. Clearly, this state can be merged
with two rounds of classical communication already in the single-shot
scenario. The proof that the state cannot be merged via one-way LOCC
even asymptotically is strongly based on the present framework of
information concentration, and the details are provided in section
7 of the Supplemental Material \cite{epaps}.

\emph{Conclusion}.---In this Letter we introduced the concentrated
information (CI) of arbitrary tripartite quantum states, provided
upper and lower bounds to it, and an explicit expression for all tripartite
pure states in the asymptotic setting and other families of mixed
states. We also investigated LOCC quantum state merging, a variation
of the standard quantum state merging protocol where the merging procedure
is performed on mixed states via LOCC only, and proved that CI is
a faithful figure of merit for this task. We also proved that distillable
entanglement, entanglement of assistance, and quantum discord can
all be expressed as exact functions of CI, and demonstrated how the
methods developed here can be used to generalize standard quantum
state merging to mixed states, thus providing novel insights on such
communication primitive. We expect that further investigation of the
concepts developed here may lead to an operational classification
of multipartite quantum states, different from what emerges from the
notions of entanglement and other quantum correlations known today.

\emph{Acknowledgements}.---We thank Marco Piani for very helpful comments.
A. S. acknowledges financial support by the Alexander von Humboldt-Foundation,
the John Templeton Foundation, the EU (IP SIQS), the ERC (AdG OSYRIS),
and the EU-Spanish Ministry (CHISTERA DIQIP). S. L. acknowledges financial
support by the Basic Science Research Program through the National
Research Foundation of Korea funded by the Ministry of Education (NRF-2012R1A1A2003441).
G. A. acknowledges financial support by the Foundational Questions
Institute (FQXi-RFP3-1317) and the ERC StG GQCOP (Grant Agreement
No.~637352).

\bibliographystyle{apsrev4-1}
\bibliography{literature}

\cleardoublepage{}

\appendix*
\setcounter{equation}{0}
\setcounter{page}{1}

\section*{Supplemental Material}

\subsection{\label{sub:proof1}Upper bound on CI}

Here we will prove that the concentrated information is bounded above
as follows: 
\begin{equation}
{\cal I}(\rho)\leq\min\left\{ I^{A:BC}(\rho),S(\rho^{A})+E_{d}^{AB:C}(\rho)\right\} .\label{eq:upbound-1}
\end{equation}
The inequality ${\cal I}(\rho)\leq I^{A:BC}(\rho)$ was already proven
in Eq.~(\ref{eq:inequality}) of the main text, and it remains to
show the inequality ${\cal I}(\rho)\leq S(\rho^{A})+E_{d}^{AB:C}(\rho)$.
For this we will show that any final state $\sigma_{f}^{ACR}=\mathrm{Tr}_{B}[\Lambda[\sigma_{i}]]$
with initial state $\sigma_{i}=\rho^{ABC}\otimes\rho^{R}$ and an
LOCC protocol $\Lambda=\Lambda_{B\leftrightarrow CR}$ satisfies the
inequality 
\begin{equation}
I^{A:CR}(\sigma_{f}^{ACR})-S(\rho^{A})\leq E_{d}^{AB:C}(\rho^{ABC}).
\end{equation}
Noting that the LOCC protocol $\Lambda=\Lambda_{B\leftrightarrow CR}$
does not affect Alice's subsystem and using the relation $E_{d}^{AB:C}(\rho^{ABC})=E_{d}^{AB:CR}(\sigma_{i})$,
this inequality is equivalent to 
\begin{equation}
S(\sigma_{f}^{CR})-S(\sigma_{f}^{ACR})\leq E_{d}^{AB:CR}(\sigma_{i}).
\end{equation}
The latter inequality can be proven by using the fact that the distillable
entanglement is bounded below as \cite{Devetak2005} 
\begin{equation}
S(\sigma_{f}^{CR})-S(\sigma_{f}^{ACR})\leq E_{d}^{A:CR}(\sigma_{f}^{ACR}),
\end{equation}
and is nonincreasing under LOCC and under discarding subsystems: 
\begin{align}
E_{d}^{A:CR}(\sigma_{f}^{ACR}) & \leq E_{d}^{AB:CR}(\sigma_{f}^{ABCR})\leq E_{d}^{AB:CR}(\sigma_{i})
\end{align}
with $\sigma_{f}^{ABCR}=\Lambda[\sigma_{i}]$. This completes the
proof.

\subsection{\label{sub:pure}One-way CI for pure states}

In the following, we will prove that the one-way concentrated information
for a pure state $\ket{\psi}=\ket{\psi}^{ABC}$ is given by 
\begin{equation}
{\cal I}_{\rightarrow}(\ket{\psi})=S(\rho^{A})+E_{a}(\rho^{AC}),\label{eq:pure states-1}
\end{equation}
where $E_{a}$ is the entanglement of assistance.

To prove Eq.~(\ref{eq:pure states-1}), note that for any state $\rho=\rho^{ABC}$
the one-way concentrated information can be written as follows: 
\begin{equation}
{\cal I}_{\rightarrow}(\rho)=\sup_{\{M_{i}^{B}\}}I^{A:CR}\bigg(\sum_{i}p_{i}\rho_{i}^{AC}\otimes\ket{i}\bra{i}^{R}\bigg)\label{eq:one-way-1}
\end{equation}
with $p_{i}\rho_{i}^{AC}=\mathrm{Tr}_{B}[M_{i}^{B}\rho]$. We will
now show that the supremum in the above expression can be achieved
over rank-1 POVMs. For this, note that any POVM $\{M_{i}\}$ can be
refined to a rank-1 POVM $\{M_{ij}\}$ such that $\sum_{j}M_{ij}=M_{i}$.
In the next step we will show that such a refined POVM leads to a
larger mutual information, i.e., 
\begin{equation}
I^{A:CR}\left(\sum_{ij}p_{ij}\rho_{ij}^{AC}\otimes\ket{ij}\bra{ij}^{R}\right)\geq I^{A:CR}\left(\sum_{i}p_{i}\rho_{i}^{AC}\otimes\ket{i}\bra{i}^{R}\right)
\end{equation}
with $p_{ij}\rho_{ij}^{AC}=\mathrm{Tr}_{B}[M_{ij}^{B}\rho]$. This
can be seen by rewriting this expression as follows: 
\begin{equation}
\sum_{ij}p_{ij}\left[S(\rho_{ij}^{AC})-S(\rho_{ij}^{C})\right]\leq\sum_{i}p_{i}\left[S(\rho_{i}^{AC})-S(\rho_{i}^{C})\right].
\end{equation}
The latter inequality is true due to the concavity of the conditional
entropy \cite{Nielsen2000}. This proves that the supremum in Eq.~(\ref{eq:one-way-1})
can be performed over rank-1 POVMs.

In the next step it is crucial to note that, for a pure state $\ket{\psi}=\ket{\psi}^{ABC}$
and a rank-1 POVM $\{M_{i}^{B}\}$, each state $\rho_{i}^{AC}$ in
Eq.~(\ref{eq:one-way-1}) is also pure. This implies that the mutual
information on the right-hand side of Eq.~(\ref{eq:one-way-1}) can
be written as follows: 
\begin{equation}
I^{A:CR}\bigg(\sum_{i}p_{i}\ket{\psi_{i}}\bra{\psi_{i}}^{AC}\otimes\ket{i}\bra{i}^{R}\bigg)=S(\rho^{A})+\sum_{i}p_{i}S(\rho_{i}^{C}).\label{eq:one-way-2}
\end{equation}
Note that the maximization over rank-1 POVMs in Eq.~(\ref{eq:one-way-1})
is equivalent to a maximization over all decompositions of the state
$\rho^{AC}$ in Eq.~(\ref{eq:one-way-2}). This completes the proof
of Eq.~(\ref{eq:pure states-1}).

\subsection{\label{sub:Asymptotic-1}Asymptotic CI for pure states}

Here we will show that for any pure state $\ket{\psi}=\ket{\psi}^{ABC}$
the regularized CI and its one-way version coincide, and both are
given as 
\begin{equation}
{\cal I}^{\infty}(\ket{\psi})={\cal I}_{\rightarrow}^{\infty}(\ket{\psi})=S(\rho^{A})+\min\{S(\rho^{A}),S(\rho^{C})\}.
\end{equation}
For this, we use Eq.~(\ref{eq:pure states}) in the main text, which
implies that the regularized one-way CI can be written as follows:
\begin{equation}
{\cal I}_{\rightarrow}^{\infty}(\ket{\psi})=S(\rho^{A})+E_{a}^{\infty}(\rho^{AC})
\end{equation}
with the regularized entanglement of assistance $E_{a}^{\infty}(\rho)=\lim_{n\rightarrow\infty}E_{a}(\rho^{\otimes n})/n$.
In the next step we use the fact that the regularized entanglement
of assistance of the state $\rho^{AC}$ is equal to the smallest of
the local entropies \cite{Smolin2005}: 
\begin{equation}
E_{a}^{\infty}(\rho^{AC})=\min\{S(\rho^{A}),S(\rho^{C})\}.
\end{equation}
It follows that for pure states the regularized one-way CI is given
by 
\begin{equation}
{\cal I}_{\rightarrow}^{\infty}(\ket{\psi})=S(\rho^{A})+\min\{S(\rho^{A}),S(\rho^{C})\}.
\end{equation}
On the other hand, recall that Theorem \ref{thm:1} in the main text
also holds in the asymptotic scenario. Applying Theorem \ref{thm:1}
to a pure state, we find that the regularized CI is bounded above
as follows: 
\begin{equation}
{\cal I}^{\infty}(\ket{\psi})\leq S(\rho^{A})+\min\{S(\rho^{A}),S(\rho^{C})\}.
\end{equation}
Combining these results completes the proof.

\subsection{\label{sub:DistillableEntanglement}CI and distillable entanglement}

In the following we consider tripartite states of the form 
\begin{equation}
\rho=\ket{\psi}\bra{\psi}^{AB_{1}}\otimes\rho^{B_{2}C},\label{eq:rho}
\end{equation}
assuming without loss of generality that all subsystems $A$, $B_{1}$,
$B_{2}$, and $C$ have the same dimension. We will prove that the
regularized CI for this set of states can be given as follows: 
\begin{equation}
{\cal I}^{\infty}(\rho)=S(\rho^{A})+\min\{S(\rho^{A}),E_{d}(\rho^{B_{2}C})\}.\label{eq:asymptotic-2}
\end{equation}

To prove this statement, we first invoke Theorem \ref{thm:1} in the
main text, which implies that the regularized CI in this case is bounded
above as follows: 
\begin{equation}
{\cal I}^{\infty}(\rho)\leq S(\rho^{A})+\min\{S(\rho^{A}),E_{d}(\rho^{B_{2}C})\}.
\end{equation}
To complete the proof, we will now show that ${\cal I}^{\infty}$
is also bounded below by the same expression. To show this, suppose
that the distillable entanglement between Bob and Charlie is smaller
than the entropy of Alice's state: 
\begin{equation}
E_{d}(\rho^{B_{2}C})=\alpha S(\rho^{A})\label{eq:Ed}
\end{equation}
with $\alpha<1$. Bob can then asymptotically teleport the fraction
$\alpha$ of his states to Charlie by using Schumacher compression
\cite{Schumacher1995}. This means that for any $\varepsilon>0$ there
exist integers $m\leq\alpha n$ and an LOCC protocol $\Lambda$ between
Bob and Charlie such that 
\begin{equation}
||\Lambda(\rho^{\otimes n})-\ket{\psi}\bra{\psi}_{ac}^{\otimes m}\otimes\ket{\psi}\bra{\psi}_{ab}^{\otimes n-m}||\leq\varepsilon.\label{eq:teleportation}
\end{equation}
Note that the states $\ket{\psi}_{ab}$ and $\ket{\psi}_{ac}$ are
both equivalent to the state $\ket{\psi}^{AB_{1}}$ in Eq.~(\ref{eq:rho}),
but $\ket{\psi}_{ab}$ is shared by Alice and Bob while $\ket{\psi}_{ac}$
is shared by Alice and Charlie. Here $||M||=\mathrm{Tr}\sqrt{M^{\dagger}M}$
is the trace norm of the operator $M$, and the integers $n$ and
$m$ satisfy the inequality 
\begin{equation}
\alpha\geq\frac{m}{n}\geq\alpha-\varepsilon.\label{eq:integers}
\end{equation}

In the next step, observe that by LOCC Bob and Charlie can transform
the pure state $\ket{\psi}_{ab}$ shared by Alice and Bob into a mixed
state $\sigma_{ac}$ shared by Alice and Charlie. It is crucial to
note that this procedure can be performed in such a way that the mutual
information of $\sigma_{ac}$ is equal to $S(\rho^{A})$ \cite{Streltsov2013}.
Together with Eq.~(\ref{eq:teleportation}) this reasoning implies
that for any $\varepsilon>0$ there exist integers $n$ and $m$ satisfying
Eq.~(\ref{eq:integers}) and an LOCC protocol $\Lambda$ between
Bob and Charlie such that 
\begin{equation}
||\Lambda(\rho^{\otimes n})-\ket{\psi}\bra{\psi}_{ac}^{\otimes m}\otimes\sigma_{ac}^{\otimes n-m}||\leq\varepsilon.
\end{equation}
With the mutual information of $\ket{\psi}_{ac}$ and $\sigma_{ac}$
given by $2S(\rho^{A})$ and $S(\rho^{A})$ respectively, it is easy
to verify that the mutual information of $\ket{\psi}\bra{\psi}_{ac}^{\otimes m}\otimes\sigma_{ac}^{\otimes n-m}$
is given by $(n+m)S(\rho^{A})$.

To complete the proof of Eq.~(\ref{eq:asymptotic-2}) we will use
continuity of the mutual information. In particular, for two states
$\rho=\rho^{XY}$ and $\sigma=\sigma^{XY}$ with $||\rho-\sigma||\leq1$,
the mutual information satisfies the following inequality: 
\begin{align}
|I^{X:Y}(\rho)-I^{X:Y}(\sigma)| & \leq3T\log_{2}d+3h(T).\label{eq:continuity-1}
\end{align}
Here, $d$ is the dimension of the total Hilbert space, $T=||\rho-\sigma||/2$
is the trace distance between $\rho$ and $\sigma$, and $h(x)=-x\log_{2}x-(1-x)\log_{2}(1-x)$
is the binary entropy. The aforementioned inequality~(\ref{eq:continuity-1})
can be proven directly by using continuity of the von Neumann entropy
\cite{Audenaert2007}.

It follows that for any $0<\varepsilon\leq1/2$ there exist integers
$n$ and $m$ satisfying Eq.~(\ref{eq:integers}) such that 
\begin{equation}
{\cal I}(\rho^{\otimes n})\geq(n+m)S(\rho^{A})-3\varepsilon\log_{2}d^{n}-3h(\varepsilon).
\end{equation}
Note that in the above expression $d$ is the dimension of a single
copy of the state $\rho$ (i.e., the dimension of $\rho^{\otimes n}$
is given by $d^{n}$). In the next step we will use Eq.~(\ref{eq:integers}),
thus arriving at the following inequality: 
\begin{align}
\frac{1}{n}{\cal I}(\rho^{\otimes n}) & \geq\left(1+\alpha\right)S(\rho^{A})-\varepsilon\left[S(\rho^{A})+3\log_{2}d^{n}\right]-3h(\varepsilon)
\end{align}
which is true for any $0<\varepsilon\leq1/2$ and some integer $n$.
By using Eq.~(\ref{eq:Ed}) this result implies that the regularized
CI is bounded below by 
\begin{equation}
{\cal I}^{\infty}(\rho)\geq S(\rho^{A})+E_{d}(\rho^{B_{2}C}),
\end{equation}
which completes the proof of Eq.~(\ref{eq:asymptotic-2}) for the
case $E_{d}(\rho^{B_{2}C})<S(\rho^{A})$. On the other hand, if $E_{d}(\rho^{B_{2}C})\geq S(\rho^{A})$,
Bob can asymptotically teleport all copies of his state to Charlie.
Using the same lines of reasoning as above we see that the regularized
CI is given by $2S(\rho^{A})$ in this case. This completes the proof
of Eq.~(\ref{eq:asymptotic-2}).

\subsection{\label{sub:CI-and-fidelity}CI and fidelity of LQSM}

Here we will prove that the fidelity of LQSM for any state $\rho=\rho^{ABC}$
is bounded below as 
\begin{equation}
{\cal F}(\rho)\geq2^{-\frac{1}{2}[I^{A:BC}(\rho)-{\cal I}(\rho)]}.
\end{equation}
In the first step, note that every LOCC operation between Bob and
Charlie can be regarded as a unitary by introducing an environment
$E$, i.e., the final state $\sigma_{f}^{ACR}$ shared by Alice and
Charlie can be written as $\sigma_{f}^{ACR}=\mathrm{Tr}_{BE}[\sigma_{f}^{ABCRE}]$
with the total final state 
\begin{equation}
\sigma_{f}^{ABCRE}=U(\rho^{ABC}\otimes\rho^{R}\otimes\rho^{E})U^{\dagger}
\end{equation}
and a unitary $U$ acting on all systems but $A$.

In the next step, we will use the main result from a recent work \cite{Fawzi2014}:
there exists a quantum operation $\Lambda$ from $CR$ to $BCRE$
which allows to recover the total final state $\sigma_{f}^{ABCRE}$
from the reduced state $\sigma_{f}^{ACR}$, with fidelity bounded
below as 
\begin{equation}
F\left(\sigma_{f}^{ABCRE},\Lambda[\sigma_{f}^{ACR}]\right)\geq2^{-\frac{1}{2}I(A:BE|CR)_{\sigma_{f}^{ABCRE}}},\label{eq:fawzi}
\end{equation}
where $I(X$:$Y|Z)$ is the conditional mutual information.

This means that by acting on his particles $C$ and $R$, Charlie
can locally recover the total final state $\sigma_{f}^{ABCRE}$ from
the reduced state $\sigma_{f}^{ACR}$ with fidelity bounded below
by Eq.~(\ref{eq:fawzi}). As the fidelity is invariant under unitaries
and nondecreasing under discarding subsystems, Charlie can obtain
the desired target state $\sigma_{t}^{ACR}$ with fidelity bounded
below by Eq.~(\ref{eq:fawzi}), and thus 
\[
{\cal F}(\rho)\geq2^{-\frac{1}{2}I(A:BE|CR)_{\sigma_{f}^{ABCRE}}}
\]
for any final state $\sigma_{f}^{ABCRE}$. The proof is complete by
noting that the conditional mutual information of $\sigma_{f}^{ABCRE}$
can be written as 
\begin{equation}
I(A:BE|CR)_{\sigma_{f}^{ABCRE}}=I^{A:BC}(\rho^{ABC})-I^{A:CR}(\sigma_{f}^{ACR}),
\end{equation}
which can be verified by inspection.

\subsection{\label{sub:Asymptotic}Asymptotic LQSM and regularized CI}

Here we will prove that perfect asymptotic LQSM implies perfect asymptotic
information concentration, i.e., 
\begin{equation}
{\cal F}^{\infty}(\rho)=1\Rightarrow{\cal I}^{\infty}(\rho)=I^{A:BC}(\rho).\label{eq:asymptotic-3}
\end{equation}
For proving this statement, it is important to note that ${\cal F}^{\infty}(\rho)=1$
implies that for any $\varepsilon>0$ there exists an integer $n\geq1$
and an LOCC protocol $\Lambda$ between Bob and Charlie such that
\begin{equation}
||\sigma_{f}-\sigma_{t}^{\otimes n}||\leq\varepsilon.\label{eq:perfect}
\end{equation}
Here, the target state $\sigma_{t}=\sigma_{t}^{ACR}$ is equivalent
to the state $\rho=\rho^{ABC}$ up to relabeling the particles $B$
and $R$, the final state $\sigma_{f}$ is given by $\sigma_{f}=\mathrm{Tr}_{B}[\Lambda[\sigma_{i}^{\otimes n}]]$
with initial state $\sigma_{i}=\rho^{ABC}\otimes\rho^{R}$, and $||M||=\mathrm{Tr}\sqrt{M^{\dagger}M}$
is the trace norm of the operator $M$.

We will now complete the proof of Eq.~(\ref{eq:asymptotic-3}) by
using Eqs.~(\ref{eq:perfect}) and continuity of mutual information
in Eq.~(\ref{eq:continuity-1}). In particular, these results imply
that for any $0<\varepsilon\leq1/2$ there exists an integer $n\geq1$
such that 
\begin{equation}
{\cal I}(\rho^{\otimes n})\geq I^{A:BC}(\rho^{\otimes n})-3\varepsilon\log_{2}d^{n}-3h(\varepsilon),
\end{equation}
where now $d$ is the dimension of the total system $ABC$. This result
implies that for states $\rho$ satisfying ${\cal F}^{\infty}(\rho)=1$
the regularized concentrated information ${\cal I}^{\infty}$ is bounded
below by $I^{A:BC}$. Noting that ${\cal I}^{\infty}$ is bounded
above by the same quantity completes the proof.

\subsection{\label{sub:LOCC}General LOCC versus one-way LOCC}

Here we will consider the following family of states 
\begin{align}
\rho & =\frac{1}{4}\left(\ket{0}\bra{0}^{B}\otimes\ket{00}\bra{00}^{AC}+\ket{1}\bra{1}^{B}\otimes\ket{10}\bra{10}^{AC}\right.\label{eq:rho-1}\\
 & \left.+\ket{\psi}\bra{\psi}^{B}\otimes\ket{01}\bra{01}^{AC}+\ket{\psi_{\bot}}\bra{\psi_{\bot}}^{B}\otimes\ket{11}\bra{11}^{AC}\right),\nonumber 
\end{align}
with orthogonal states $\ket{\psi}$ and $\ket{\psi_{\bot}}$ such
that $0<|\braket{0|\psi}|<1$. In particular, we will prove that the
state cannot be merged via one-way LOCC even in the asymptotic scenario.
Using the same arguments as in section \ref{sub:Asymptotic} of this
Supplemental Material, it is enough to prove that the regularized
one-way concentrated information is strictly below the total mutual
information: 
\begin{equation}
{\cal I}_{\rightarrow}^{\infty}(\rho)<I^{A:BC}(\rho).\label{eq:proof_theorem2-1}
\end{equation}

For proving this statement, we will use results from section \ref{sub:Bound}
of this Supplemental Material, which imply that a state $\rho=\rho^{ABC}$
cannot be merged via one-way LOCC in the asymptotic scenario if its
regularized discord is strictly larger than the mutual information
of $\rho^{BC}$ 
\begin{equation}
\lim_{n\rightarrow\infty}\frac{1}{n}\delta^{AC|B}(\rho^{\otimes n})>I^{B:C}(\rho^{BC}).
\end{equation}
Using the fact that for the family of states considered here the reduced
state $\rho^{BC}$ is maximally mixed, it remains to show that the
regularized discord is nonzero. This can be seen by noting that the
state given in Eq.~(\ref{eq:rho-1}) has nonzero discord $\delta^{AC|B}(\rho)>0$,
and by proving that the discord is additive for this family of states.
Additivity is shown in section \ref{sub:Additivity} of this Supplemental
Material, and the proof is complete.

\subsection{\label{sub:Bound}Bounding one-way CI}

In this section we will prove the following bound for the one-way
CI: 
\begin{equation}
{\cal I}_{\rightarrow}\left(\rho^{ABC}\right)\leq I^{A:BC}\left(\rho^{ABC}\right)+I^{B:C}\left(\rho^{BC}\right)-\delta^{AC|B}\left(\rho^{ABC}\right),\label{eq:bound}
\end{equation}
where $\delta$ is the quantum discord.

For proving this, we first note that the amount of discord $\delta^{X|Y}$
in a quantum state $\rho^{XY}$ can be expressed as follows \cite{Streltsov2013,Streltsov2014}:
\[
\delta^{X|Y}\left(\rho^{XY}\right)=I^{X:Y}\left(\rho^{XY}\right)-\sup_{\left\{ M_{i}^{Y}\right\} }I^{X:Z}\left(\sum_{i}\mathrm{Tr}_{Y}\left[M_{i}^{Y}\rho^{XY}\right]\otimes\ket{i}\bra{i}^{Z}\right),
\]
and the supremum is taken over all POVMs $\{M_{i}^{Y}\}$ on the subsystem
$Y$. For proving the inequality (\ref{eq:bound}) we will use the
fact that its right-hand side can also be written as $I^{A:C}(\rho^{AC})+I^{B:AC}(\rho^{ABC})-\delta^{AC|B}(\rho^{ABC})$
which is further equivalent to $I^{A:C}(\sigma^{AC})+\sup_{\{M_{i}^{B}\}}I^{AC:R}(\sigma^{ACR})$
with supremum over all POVMs $\{M_{i}^{B}\}$ and the state $\sigma^{ACR}$
is defined as 
\begin{equation}
\sigma^{ACR}=\sum_{i}\mathrm{Tr}_{B}\left[M_{i}^{B}\rho^{ABC}\right]\otimes\ket{i}\bra{i}^{R}.
\end{equation}
Note that the same state $\sigma^{ACR}$ can also be used to express
the one-way concentrated information as follows: 
\begin{equation}
{\cal I}_{\rightarrow}(\rho^{ABC})=\sup_{\left\{ M_{i}^{B}\right\} }I^{A:CR}\left(\sigma^{ACR}\right).
\end{equation}
Using the aforementioned results, the inequality (\ref{eq:bound})
is equivalent to 
\begin{equation}
\sup_{\left\{ M_{i}^{B}\right\} }I^{A:CR}\left(\sigma^{ACR}\right)\leq I^{A:C}\left(\sigma^{AC}\right)+\sup_{\left\{ M_{i}^{B}\right\} }I^{AC:R}\left(\sigma^{ACR}\right).
\end{equation}
The proof of Eq.~(\ref{eq:bound}) now follows by using the inequality
\begin{equation}
I^{A:CR}\left(\sigma^{ACR}\right)\leq I^{A:C}\left(\sigma^{AC}\right)+I^{AC:R}\left(\sigma^{ACR}\right),
\end{equation}
which is a direct consequence of the subadditivity of von Neumann
entropy \cite{Nielsen2000}.

\subsection{\label{sub:Additivity}Additivity of discord}

Here we will show that discord $\delta^{X|Y}$ is additive for states
of the form 
\begin{equation}
\rho^{XY}=\sum_{i}p_{i}\ket{i}\bra{i}^{X}\otimes\ket{\psi_{i}}\bra{\psi_{i}}^{Y}.\label{eq:additivity}
\end{equation}
This can be seen by using the Koashi-Winter relation, relating quantum
discord $\delta$ and entanglement of formation $E_{f}$ as follows
\cite{Koashi2004,Fanchini2011,Madhok2011,Cavalcanti2011}: 
\begin{equation}
\delta^{X|Y}(\rho^{XY})=E_{f}(\rho^{XZ})-S(\rho^{XY})+S(\rho^{Y}),
\end{equation}
where the total system $XYZ$ is in a pure state $\ket{\psi}^{XYZ}$.
In the next step, note that the state $\rho^{XY}$ in Eq.~(\ref{eq:additivity})
can be purified as follows: 
\begin{equation}
\ket{\psi}^{XYZ}=\sum_{i}\sqrt{p_{i}}\ket{i}^{X}\otimes\ket{\psi_{i}}^{Y}\otimes\ket{i}^{Z}.
\end{equation}
In the final step, we note that the state $\rho^{XZ}$ has the following
form: 
\begin{equation}
\rho^{XZ}=\sum_{i,j}a_{ij}\ket{i}\bra{j}^{X}\otimes\ket{i}\bra{j}^{Z}
\end{equation}
with parameters $a_{ij}=\sqrt{p_{i}p_{j}}\braket{\psi_{j}|\psi_{i}}$.
States of this form are also known in the literature as maximally
correlated states, and their entanglement of formation is known to
be additive \cite{Horodecki2003}. This also implies that quantum
discord is additive for the states given in Eq.~(\ref{eq:additivity}). 
\end{document}